\documentclass[12pt,aps,a4paper, floats,nofootinbib,amssymb,superscriptaddress]{revtex4}

\usepackage[sort&compress]{natbib}
\usepackage{epsf,epsfig}
\usepackage{amsmath}
\usepackage{hyperref}\usepackage{subfigure}
\usepackage{graphicx}
\usepackage{color}
\usepackage{mathrsfs}
\usepackage{enumitem}

\newcommand{\be}{\begin{equation}}
\newcommand{\ee}{\end{equation}}
\newcommand{\bea}{\begin{eqnarray}}
\newcommand{\eea}{\end{eqnarray}}

\newcommand{\ba}{\begin{array}}
\newcommand{\ea}{\end{array}}
\newcommand{\bi}{\begin{itemize}}
\newcommand{\ei}{\end{itemize}}

\newcommand{\redcross}{\color{red}-}
\newcommand{\doublepole}{{\color{red}{\bigotimes}}}
\newcommand{\doublerow}[2]{\hspace{0.3cm} \begin{matrix} #1 \\ #2 \end{matrix} \hspace{0.3cm}}

\newcommand{\ih}{-\frac{i}{2}}

\makeatletter

\begin{document}


\title{\vspace*{1.in}
 Accommodating scalar resonances in the HEFT
\vspace*{0.5cm}
}

\author{I\~nigo  Asi\'ain}\email{iasiain@icc.ub.edu}\affiliation{Departament de F\'isica Qu\`antica i Astrof\'isica\,,
Institut de Ci\`encies del Cosmos (ICCUB), \\
Universitat de Barcelona, Mart\'i Franqu\`es 1, 08028 Barcelona, Spain}
\author{Dom\`enec Espriu}\email{espriu@icc.ub.edu}
\affiliation{Departament de F\'isica Qu\`antica i Astrof\'isica\,,
Institut de Ci\`encies del Cosmos (ICCUB), \\
Universitat de Barcelona, Mart\'i Franqu\`es 1, 08028 Barcelona, Spain}
\author{Federico Mescia}\email{mescia@ub.edu}\affiliation{Departament de F\'isica Qu\`antica i Astrof\'isica\,,
Institut de Ci\`encies del Cosmos (ICCUB), \\
Universitat de Barcelona, Mart\'i Franqu\`es 1, 08028 Barcelona, Spain}\affiliation{Istituto Nazionale di Fisica Nucleare, Laboratori Nazionali di Frascati, \\ C.P.~13, 00044 Frascati, Italy}

\begin{abstract}
\vspace*{0.7cm}
\noindent
Loss of unitarity in an effective field theory is often cured by the appearance of dynamical resonances, revealing the presence of new degrees
  of freedom. These resonances may manifest themselves when suitable unitarization techniques are implemented in the effective theory, which in
  the scalar-isoscalar channel require using the coupled-channel formalism. Experimental detection of a resonance would  provide precious information
  on the couplings and constants of the relevant effective theory. Conversely, the absence of a resonance where the unitarized effective theory predicts
  it should be allows us to rule out a certaing range of couplings that would otherwise be allowed. Likewise, the appearence of unphysical (e.g. acausal)
  resonances is telling us that no UV completion could give rise to the corresponding couplings in the effective theory. In this talk we summarize 
  the systematical procedure we have implemented in order to confront the effective theory with the absence or presence of resonances
  in the vector boson fusion channel at the LHC.
\end{abstract}

\maketitle

\newpage
\thispagestyle{empty}
\tableofcontents

\thispagestyle{empty}
\newpage
 \setcounter{page}{0}

\section{Introduction}
A clear indication of the existence of New Physics (NP) with strong interactions beyond-the-standard model (BSM) would be the
appearence of unexpected resonances in the spectrum. A promising place to look for such resonances is in the  elctroweak symmetry breaking
sector (EWSBS), consisting of the gauge bosons, their associated Goldstones and the Higgs boson, responsible for the generation of the masses of
the elementary particles of the Standard Model (SM). While the general ideas behind the spontaneous
symmetry breaking (SSB) mechanism seems well established, the nature of the Higgs and its interactions are not so obvious. In particular, the
choice of the potential, built upon the principles of renormalizability and simplicity, is largely untested.

It is natural to look for NP in the EWSBS at the LHC is in vector boson fusion (VBF), and particularly in the interactions of the longitudinal components
of the gauge bosons, as these are the ones intimately related to the SSB of the vacuum.
This is perhaps best understood in the framework of the  Equivalence Theorem (ET)\cite{Cornwall:1974km}, which states that at high energies compared
to the  electroweak scale, the longitudinal component of the gauge bosons can be susbstituted by their associated Goldstones.
A throughout discussion of the ET and the error that one assumes with its usage can be found in Refs.~\cite{Espriu:1994ep,paper2_doi}.

Here we summarize a systematical procedure to derive interesting phenomenological bounds on the coefficients of the Higgs potential, based on the properties
of possible scalar resonances emerging in the context of a strongly interacting EWSBS.

\section{The HEFT and the chiral parameter space}
The Higgs effective field theory (HEFT) is a non-linear chiral Lagrangian describing the electroweak interactions up to the TeV scale
that contains the Higgs boson as an $SU(2)$ singlet, in clear contrast to the linear case. The three Goldstones $\omega^a$ are included in a
matrix exponential $U=\exp\left(i\omega^a\sigma^a/v\right)$ taking values in the coset $SU(2)_L\times SU(2)_R/SU(2)_V$.

We restrict ourselves to operators that respect the custodial symmetry, a limit in which the gauge bosons transform exactly as a triplet under
the so-called custodial group. In our case, this limit is obtained by setting $g^\prime=0$, where $g^\prime$ is the gauge coupling associated to
the abelian hypercharge group of the SM. This approximation seems justified by the $\rho$ parameter value \cite{ROSS1975135}.

Up to next-to-leading-order (NLO) we have the Lagrangian density
\begin{equation}\label{eq: lag2}
\begin{split}
  \mathcal{L}_2 =&-\frac{1}{2g^2}\text{Tr}\left(\hat{W}_{\mu\nu}\hat{W}^{\mu\nu}\right)-
\frac{1}{2g^{\prime 2}}\text{Tr}\left(\hat{B}_{\mu\nu}\hat{B}^{\mu\nu}\right)
  +\frac{v^2}{4}\mathcal{F}(h)\text{Tr}\left(D^{\mu}U^{\dagger}D_{\mu}U\right)+\frac{1}{2}\partial_{\mu}h\partial^{\mu}h \\ 
  &-V(h) \vspace{0.2cm} \\
\end{split}
\end{equation}
\begin{equation}\label{eq: lag4}
\begin{split}
  \mathcal{L}_4 =&-i a_3\text{Tr}\left(\hat{W}_{\mu\nu}\left[V^{\mu},V^{\nu}\right]\right)
  +a_4 \left(\text{Tr}\left(V_{\mu}V_{\nu}\right)\right)^2
  +a_5 \left(\text{Tr}\left(V_{\mu}V^{\mu}\right)\right)^2+\frac{\gamma}{v^4}\left(\partial_{\mu}h\partial^{\mu}h\right)^2\\
  &+\frac{\delta}{v^2}\left(\partial_{\mu}h\partial^{\mu}h\right)\text{Tr}\left(D_{\mu}U^{\dagger}D^{\mu}U\right)
  +\frac{\eta}{v^2}\left(\partial_{\mu}h\partial_{\nu}h\right)\text{Tr}\left(D^{\mu}U^{\dagger}D^{\nu}U\right)\\
&+i\chi\,\text{Tr}\left(\hat{W}_{\mu\nu}V^{\mu}\right)\partial^{\nu}\mathcal{G}(h)
\end{split}
\end{equation}
with the usual vector structures $V_\mu=D_\mu U^\dagger U$, $\hat{W}_{\mu\nu}=\frac{1}{2}gW_\mu^a\sigma^a$, the $\mathcal{ F}, \mathcal{G}$ flare-functions and the
parametrization of the Higgs potential
\begin{equation}\label{eq: higgs_potential}
\begin{split}
&\mathcal{F}(h)=1+2a\frac{h}{v}+b\left(\frac{h}{v}\right)^2+o\left(h/v\right)^3, \quad \mathcal{G}(h)=1+b_1\frac{h}{v}+o\left(h/v\right)^2\\
&V(h)=\frac{1}{2}M_h^2h^2+d_3\lambda_{SM}vh^3+d_4\frac{\lambda_{SM}}{4}h^4+\cdots.
\end{split}
\end{equation}

All operators are sorted by the number of derivatives and soft mass scales. More information about this so-called \textit{chiral counting}
can be found in Ref.~\cite{Asiain:2021lch}. The coefficients accompanying these operators, in general different to their SM values, are referred to
as \textit{anomalous} couplings. One can easily recover the SM by setting all the anomalous couplings in $\mathcal{L}_2$ to 1,
and the ones in $\mathcal{L}_4$ to zero, as they are absent in the SM. Table \ref{tab_ET_chiralparams} collects all the relevant experimental
bounds for the anomalous couplings up to date. The ones absent are not experimentally constrained at the moment.

\begin{table}[tb]
\begin{center}
\renewcommand{\arraystretch}{1.2}
\begin{tabular}{|c|c|c| }
\hline
Couplings & Ref. & Experiments\\ 
\hline \hline
$0.89<a<1.13$  & \cite{deBlas:2018tjm}
& CMS (from $H\to VV$)  \\ \hline 
$0.55<b<1.49$ & \cite{CMS:2024fkb} & ATLAS (from $HH\to X$) \\ \hline
$-0.4<d_3<6.3$ &\cite{ATLAS:2022jtk} & ATLAS (from $H\to X$ and $HH\to X$) \\ \hline
 $-0.0061<a_4 < 0.0063$&\cite{CMS:2019uys} & CMS (from $WZ\to 4l$) \\ \hline
$-0.0094<a_5 < 0.0098$ &  \cite{Sirunyan:2019der} & CMS (from $WZ/WW\to 2l2j$) \\  \hline
\end{tabular}
\caption{{\small
    Current experimental constraints on bosonic HEFT anomalous
    couplings at 95\% CL. $X$ stands for different combinations of $l^{+}l^{-}$, $b\bar{b}$ and $\gamma\gamma$ that can participate in the process of Higgs decays.}} \label{tab_ET_chiralparams}
\end{center}
\end{table}

With $\mathcal{L}_2+\mathcal{L}_4$, we obtain all the relevant $2\to 2$ amplitudes $W_LW_L\to W_LW_L$, $W_LW_L\to hh$ and $hh\to hh$. Each of these
three amplitudes up to NLO are decomposed in a tree-level contribution, obtained from $\mathcal{L}_2+\mathcal{L}_4$, and a one-loop contribution
that for simplicity is derived using the ET containing only $\mathcal{L}_2$ interactions. The ultraviolet divergences from the one-loop calculation
are absorbed by the proper redefinitions of the parameters of the Lagrangian. Detailed information about amplitudes
and local counterterms is provided in Ref.~\cite{Asiain:2021lch}.

\section{Unitarized partial waves}
The expansion in derivatives (momenta) typically leads to amplitudes that quickly violate unitarity, even for small departures from the SM values.
In order to avoid this unphysical behavior that would irredemiably lead to an overestimation of any  NP signal, one must \textit{unitarize} the
amplitudes by means of some unitarization technique. Most of them are built in the language of  partial-waves, where the unitarity condition
acquires simple expressions. We are interested in the projection of the amplitudes into a specific channel with fixed
isospin and spin ($IJ$). The isoscalar-scalar waves are obtained using
\begin{equation}
t_{00}^{(n)}=\frac{1}{64\pi}\int_1^1d\left(\cos\theta\right)\,T_0^{(n)}(s,\cos\theta),
\end{equation}
where $T_0^{(n)}$ is the amplitude with $I=0$ for each of the $2\to 2$ amplitudes at chiral order $n$, see Ref.~\cite{paper2_doi}.

The Inverse Amplitude Mehtod (IAM), extensively used in low energy QCD, is the one chosen:
\begin{equation}\label{eq: tIAM}
t_{IJ}\approx t^{(2)}_{IJ}+t^{(4)}_{IJ}+\cdots,\qquad t_{IJ}^{\text{IAM}}=t^{(2)}_{IJ}\left(t_{IJ}^{(2)}-t_{IJ}^{(4)}\right)^{-1}t^{(2)}_{IJ}.
\end{equation}

For the case $IJ=00$, Eq.~(\ref{eq: tIAM}) is applied in a matrix version containing all the possible processes that can participate
in the channel at chiral order $n$:
\begin{equation}\label{eq: t00_matrix}
t_{00}^{(n)}=\begin{pmatrix}t_{00}^{WW,(n)} & t_{00}^{Wh, (n)} \\ t_{00}^{Wh, (n)} & t_{00}^{hh, (n)}\end{pmatrix},
\end{equation}
where $WW$, $Wh$ and $hh$ indicate $W_LW_L\to W_LW_L$, $W_LW_L\to hh$ and $hh\to hh$, respectively. As all these amplitudes mix along the unitarization
process when the Eq.~(\ref{eq: tIAM}) is applied, this  method recieves the name of \text{coupled-channel} formalism.

A scalar resonance, if present, appears in the spectrum as a pole of the unitarized wave $t^{IAM}_{00}$. Looking at Eq.~(\ref{eq: tIAM}), this occurs
when $\det(t^{(2)}_{00}(s_R)-t^{(4)}_{00}(s_R))=0$ for a complex value of the Mandelstam variable $s_R=M_S^2-\frac{i}{2}\Gamma_S M_S$,
where $M_S$ and $\Gamma_S$ define the mass and width of the resonance. A resonance is considered physical when it appears on the second Riemann
sheet of the complex $s$-plane, after analytical continuation across the physical
cut in the real axis. Additionally, it must satisfy the condition $\Gamma<M/4$. Whenever an analytical continuation is not feasible due to the complexity
of the amplitudes, the way we choose to determine whether a resonance is physical or not is by the phase-shift criterion: the phase $\delta$
(in the complex sense) of an amplitude that contains a physical resonance presents a shift from $\pi/2$ to $-\pi/2$ at the real pole position.
A shift in the opposite direction is considered unphysical, due to $\Gamma^{-1}\sim \partial\delta (s)/\partial \sqrt{s}$.
\section{Summary of the results}
For this analysis we assume that any set of anomalous couplings leading to spurious resonances lacks a proper UV completion and cannot define a valid HEFT.
Additionally, a scalar resonance lighter than 1.8 TeV should have already been observed, allowing us to rule out the corresponding set of anomalous couplings.
This threshold is motivated by the study in Ref.~\cite{Rosell:2020iub}, which constrains possible vector masses to $M_V \gtrsim 2$ TeV. Since experience
suggests that scalar multiplets tend to be lighter than their vector counterparts, we adopt a slightly relaxed bound of 1.8 TeV.

Because the HEFT parameter space is large,  it helps noticing a clear hierarchy between $a_4$ and $a_5$ (that have the largest number of derivatives)
and the remaining couplings, even if nominally of the same chiral order. However it is worth checking the relevance of other operators, in
particular those involving the propagation of the transverse gauge degrees of freedom inside the loops
A detailed discussion about their relative contributions can be found in Ref.~\cite{paper2_doi}. For instance, in 
Table \ref{table: differences_arnan} we show the impact of going beyond the $g=0$ case (the naive ET limit) for some specific benchmark points (BPs)
for $a_4-a_5$ in the limit $a=b^2$. The inclusion of transverse modes translates into heavier resonances by a difference of a $2-3\%$.
\begin{table}[h!]
\begin{center}
\begin{tabular}{|c|c|c|c|c|}
\hline
    $\sqrt{s_S} \, (GeV)$ &$a_4\cdot 10^4$ & $a_5\cdot 10^4$ & $g=0$ & $g\neq 0$\\ \hline
$ $  & $ \quad 1 \quad $ &     $\quad -0.2 \quad $    & $\quad 1805 \ih 130 \quad $ & $\quad 1856 \ih 125\quad$            \\ \hline
$ $  & $\quad 2 \quad$ &   $\quad -1  \quad$      & $\quad 2065 \ih 160 \quad$ & $\quad 2119 \ih 150 \quad$\\ \hline
$ $ & $\quad 3.5 \quad $ &  $\quad -2 \quad$       & $\quad 2175\ih 170 \quad$ & $\quad 2231 \ih 163 \quad$ \\ \hline
\end{tabular}
\caption{{\small Values for the location of the scalar poles $\sqrt{s_S}=M_S-\frac{i}{2}\Gamma_S$ for $g=0$ and $g\neq 0$ for some points in the $a_4-a_5$ plane and in the decoupling limit $b=a^2$ within the nET with $a=b=1$. All other couplings are set to the SM values. Note that the coupling to other $I=0$ channels is ignored here for the purpose of assessing the effect of switching on the transverse modes.}}\label{table: differences_arnan}
\end{center}
\end{table}

However, once one moves from the case $g=0$ (i.e. the naive ET), the decupling limit does not take place even for setting $a=b^2$,
so a coupled-channel analysis is required for our study. This is shown Table \ref{table: differences_arnan}. 
\begin{table}[h!]
\begin{center}
\begin{tabular}{|c|c|c|c|c|c|}
\hline
$  $ & $ a_4\cdot 10^4 $ & $a_5\cdot 10^4$ &  $\text{S.C.} $ & $\text{C.C.} $ & $M_V\ih \Gamma_V$\\ \hline
$ \text{BP1}$ & $ 3.5  $ & $ 1 $   &  $ 1044\ih 50  $ & $ {\bf 1844\ih 487} $ & $2540\ih 27 $ \\ \hline
$ \text{BP2}$ & $ -1 $  & $ 2.5 $   &  $ 1219\ih 75 $ & ${\bf 2156\ih 637} $ & $\redcross$ \\ \hline
$ \text{BP3}$ & $ 1 $  & $1 $   &  $1269\ih 75 $ & ${\bf 2244\ih 675} $ & $\redcross$ \\ \hline
\end{tabular}
\caption{{\small Properties of the scalar resonances for the selected benchmark points in the $a_4-a_5$ plane, with the $\mathcal{O}(p^2)$ parameters set to their standard values, in both single-channel (S.C.) and coupled-channel (C.C.) formalism. We also include the values of the properties of vector resonances if present. The dots
    indicate the absence of a zero in the determinant. The $\mathcal{O}(p^2)$ chiral parameters are set to their SM values. We see that coupling channels modifies very substantially masses and widths. Poles not fulfilling the resonance condition are in boldface.}}\label{table: BPs}
\end{center}
\end{table}
One immediately notices in  Table \ref{table: BPs} is that when (correctly) considering coupled channels, the results differ considerably from the ones
obtained in single channel and the resonance masses  and widths visibly increase. Recall that here we are assuming $a=b=1$ where naively one would expect
to have decoupling (this is the case in the nET), but this is not so because  $g\neq 0$. In fact some of the would-be resonances even dissapear
as such by just becoming broad enhancements. 

With respect to the vector case, where only $a_4$ and $a_5$ matters as they basically determine the position of the vector resonances, the HEFT parameter space
to study scalar resonances is much larger, even after dropping $a_3$ and $\zeta\equiv b_1\chi$ from our analysis, as more processes are involved. 
Let us now proceed with the stydy in the cas $a=b=d_3=d_4=1$---the SM case at LO---and assuming natural values for these extra NLO couplings---they do not exceed
an absolute value of $10^{-3}$---. The following results for the position of the resonances are found for BPs where only scalar resonances---and no vector nor tensor---emerge

\begin{table}[h!]
\begin{tabular}{|c|c|c|c|c|c|}
\hline
    $M_S\ih \Gamma_S$ & $\gamma=0$  & $\gamma=0.5\cdot 10^{-4}$ & $\gamma=1\cdot 10^{-4} $ & $\gamma=-0.5\cdot 10^{-4}$ & $\gamma=-1\cdot 10^{-4}$  \\ \hline
 $ \text{BP1} $ &  $\bf 1844\ih 487 $    & $1668\ih 212 $   & $1594\ih 162 $ & $ \redcross $ &  $ \redcross$  \\ \hline
 $ \text{BP2} $ &  $\bf 2156\ih 637  $    & $ 1881\ih 212  $  & $ 1781\ih 162 $ & $\redcross $ & $\redcross $    \\ \hline
 $ \text{BP3} $ &  $ \bf 2244\ih 675 $    & $1931\ih 200   $  & $1831\ih 162  $ & $\redcross $ & $\redcross $   \\ \hline
\end{tabular}
\caption{{\small Pole position for the benchmark points in Table \ref{table: BPs} varying the $\mathcal{O}(p^4)$ parameter $\gamma$.
    The rest of the parameters are set to their SM values. Values in boldface indicate broad resonances that do not satisfy $\Gamma<M/4$.}}\label{table: differences_coupled_gamma}
\end{table}

\begin{table}[h!]
\begin{tabular}{|c|c|c|c|c|c|}
\hline
    $M_S\ih \Gamma_S$ & $\delta=0$  & $\delta=0.5\cdot 10^{-4}$ & $\delta=1\cdot 10^{-4} $ & $\delta=-0.5\cdot 10^{-4}$ & $\delta=-1\cdot 10^{-4}$  \\ \hline
 $ \text{BP1} $ &  $\bf 1844\ih 487 $    & $ 1744\ih 362$   & $1669\ih 300   $   & $\bf 1994\ih 1100  $ & $ \doublepole $    \\ \hline
 $ \text{BP2} $ &  $\bf 2156\ih 637 $    & $1981\ih 387   $  & $1869\ih 300  $   & $\bf 2644\ih \Gamma  $ & $\redcross $    \\ \hline
 $ \text{BP3} $ &  $\bf 2244\ih 675$    & $2031\ih 400 $  & $1906\ih 287   $   & $\redcross $ & $\redcross $      \\ \hline
\end{tabular}
\caption{{\small Pole position for the benchmark points in Table \ref{table: BPs} varying the $\mathcal{O}(p^4)$ parameter $\delta$.
    The rest of the parameters are set to their SM values. Values in boldface indicate broad resonances that
    do not satisfy $\Gamma<M/4$.The symbols $~\redcross$ and $\doublepole$ are introduced to represent the absence of a zero in the determinant of the
    IAM matrix and the appearence of a second pole that is non-physical following the phase-shift criteria, respectively. }}\label{table: differences_coupled_delta}
\end{table}
\begin{table}[h!]
\begin{tabular}{|c|c|c|c|c|c|}
\hline
    $M_S\ih \Gamma_S$ & $\eta=0$  & $\eta=0.5\cdot 10^{-4}$ & $\eta=1\cdot 10^{-4} $ & $\eta=-0.5\cdot 10^{-4}$ & $\eta=-1\cdot 10^{-4}$  \\ \hline
 $ \text{BP1} $ &  $\bf 1844\ih 487$    & $1806\ih 437  $  & $1769\ih 387 $ & $\bf 1881\ih 575$ & $\bf 1931\ih 712$     \\ \hline
 $ \text{BP2} $ &  $\bf 2156\ih 637$    & $2094\ih 512  $  & $2031\ih 437 $ & $\bf 2256\ih 887 $ & $\bf 2394\ih \Gamma$      \\ \hline
 $ \text{BP3} $ &  $\bf 2244\ih 675 $    & $2156\ih 537  $  & $2094\ih 450 $ & $\bf 2356\ih 925 $ & $\bf 2544\ih \Gamma $      \\ \hline
\end{tabular}
\caption{{\small Pole position for the benchmark points in Table \ref{table: BPs} varying the $\mathcal{O}(p^4)$ parameter $\eta$.
    The rest of the parameters are set to their SM values. Values in boldface indicate broad resonances that do not satisfy $\Gamma<M/4$.}}\label{table: differences_coupled_eta}
\end{table}
From  Tables \ref{table: differences_coupled_gamma}-\ref{table: differences_coupled_eta} above we can see a different scenario from the one
in the vector-isovector case. The location of the pole changes $15\%-20\%$
when we use reasonable values of $\gamma$ and $\delta$ ($\sim 10^{-4}$) and softer variations of around $4\%-8\%$ for values of $\eta$ of the same order.

\begin{figure}
\centering
\includegraphics[clip,width=7.0cm,height=6.5cm]{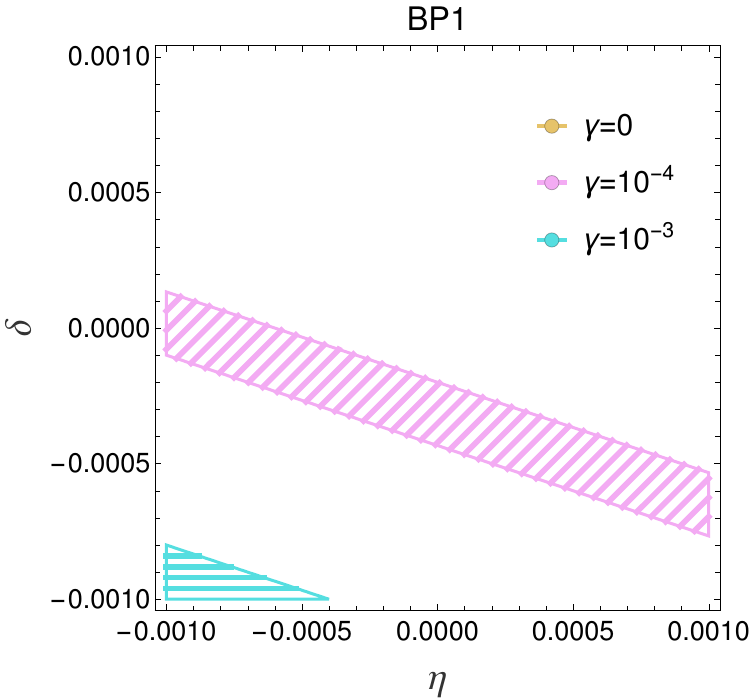} %
\includegraphics[clip,width=7.0cm,height=6.5cm]{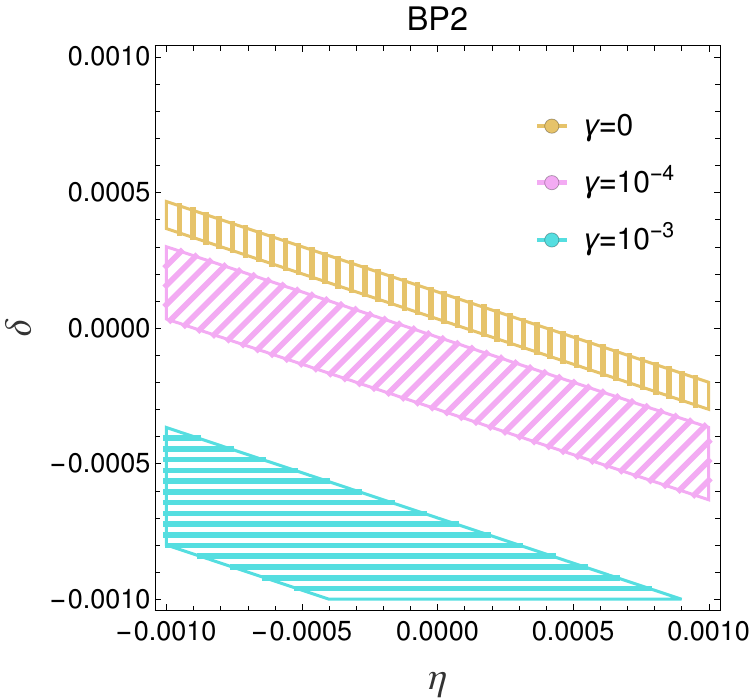} %
\caption{\small{Regions in the $\delta-\eta$ plane where physical resonances satisfying $M_S>1.8$ TeV and for two of the benchmark points in Table \ref{table: BPs} appear for different values of $\gamma$: $\gamma=0$ (golden vertical lines), $\gamma=10^{-4}$ (pink tilted lines) and $\gamma=10^{-3}$ (blue horizontal lines). For all the values of $\gamma$, the region above the bands are excluded by the presence of a non-physical pole. Below the bands we find no resonances.}}%
\label{fig: swept_gamma}
\end{figure}

However, these results could be improved by studying the more general scenario where they are all non-zero. When studying
the combined effect in the $\delta-\eta$ plane, we ontain the results in Figure \ref{fig: swept_gamma}. No matter the value of $\gamma$ or the benchmark
point selected, the presence of an unphysical pole leads us to exclude the parameter space above the bands. We also find that the greater the value of $\gamma$ is,
the more restriction we find (there are more excluded space above the band), especially for \text{BP1}. Below the bands, we find a nonresonant scenario. 

Finally, one of the more interesting results  of this systematic analysis comes from varying the couplings of the Higgs potential. In the case of the trilinear
coupling of the Higgs potential $d_3$, that now enters at tree level in the determination of the scalar resonances due to the mixing, we find that for
$d_3\gtrsim 2.5$ a second pole clearly appears (notation pole1 over pole2) in the low-energy region around $\sim 1$ TeV and it is also
physical because it is found in the second Riemann sheet of the complex $s$ plane. However, one of the physical poles is located at energy scales much
lower than our preestablished bound of $1.8$ TeV, so, in principle, the corresponding set of parameters should be discarded. The results are shown in
Table \ref{table: differences_coupled_d3_gamma}. In fact, there are already hints of this first resonance at $d_3= 1.7$. Different non-zero, but yet natural,
values for $\gamma$ do not alter the results signnificantly. For the case of $d_4$, driving modifications in the quartic Higgs self-coupling, we can repeat the same analysis to find the reuslts of Table \ref{table: differences_coupled_d4}.
\begin{table}[h!]
\centering
\resizebox{16cm}{!}{
\begin{tabular}{|c|c|c|c|c|c|c|}
\hline
    $M_S\ih \Gamma_S$ & $d_3=0.5 $ & $d_3=1$ & $d_3=2$ & $d_3=3 $ & $d_3=4 $ & $d_3=5 $  \\ \hline
  $ \text{BP1} $  &  \parbox[c][1.2cm][c]{2cm}{ $1769\ih 275 $}  & $1668\ih 212 $ & $ 1544\ih 112 $  & $\doublerow{994\ih 23}{1569\ih 25}$ & $\doublerow{1044\ih 37}{1769\ih 34}$ & $\doublerow{994\ih 27}{1994\ih 54}$  \\ \hline
    $ \text{BP2} $  &  \parbox[c][1.2cm][c]{2cm}{$1981\ih 262 $} & $1881\ih 212 $  & $1719\ih 125  $  & $\doublerow{1106\ih 27}{1656\ih 50}$ & $\doublerow{1219\ih 37}{1781\ih 34}$ & $\doublerow{1118\ih 26}{1994\ih 50}  $  \\ \hline
 $ \text{BP3} $ &  \parbox[c][1.2cm][c]{2cm}{$ 2031\ih 250$} & $1931\ih 200 $ & $1769\ih 125  $  & $\doublerow{1131\ih 37}{1681\ih 38}$ & $\doublerow{1269\ih 37}{1781\ih 27}$ & $\doublerow{1231\ih 23}{1994 \ih 53}$  \\ \hline
\end{tabular}
}
\caption{{\small Values of the pole position of the benchmark points in Table \ref{table: BPs} with $\gamma=0.5\cdot 10^{-4}$ changing $d_3$. The rest of the parameters are set to their SM values. The cells with two complex numbers indicate the pole position of the two physical Breit-Wigner poles in the denominator of the unitarized amplitude.}}\label{table: differences_coupled_d3_gamma}
\end{table}

\begin{table}[h!]
\centering
\resizebox{16cm}{!}{
\begin{tabular}{|c|c|c|c|c|c|c|c|}
\hline
    $M_S\ih \Gamma_S$ & $ d_4=0.5$ & $d_4=1$  & $d_4=2$ & $d_4=3 $ & $d_4=4 $ & $d_4=5 $ & $d_4=8 $ \\ \hline
    $ \text{BP1} $ & \parbox[c][1.2cm][c]{2cm}{$1794\ih 250 $} & $1668\ih 212  $    & $1494\ih 137 $  & $1381\ih 112 $ & $1306\ih 87  $ & $1256\ih 75 $ & $1169\ih 50 $ \\ \hline
    $ \text{BP2} $ & \parbox[c][1.2cm][c]{2cm}{$1981\ih 225 $} & $1881\ih 212  $    & $1719\ih 175 $  & $1606\ih 125  $ & $1531\ih 112  $ & $1481\ih 87 $ & $1381\ih 75 $ \\ \hline
 $ \text{BP3} $ & \parbox[c][1.2cm][c]{2cm}{$2031\ih 225 $} & $1931\ih 200  $    & $1781\ih 162 $  & $1669\ih 137  $ & $1594\ih 112  $ & $1544\ih 100 $ & $1444\ih 75$ \\ \hline
\end{tabular}
}
\caption{{\small Values of the pole position of the benchmark points in Table \ref{table: BPs} changing $d_4$ with $\gamma=0.5\cdot 10^{-4}$. The rest of the parameters are set to their SM values.}}\label{table: differences_coupled_d4}
\end{table}

From Table \ref{table: differences_coupled_d4} we can say that, if all the rest of parameters are set to their SM values, we
could exclude values of $d_4\gtrsim 2$ for \text{BP2} and \text{BP3} and \text{BP1} would be excluded since these parameters lead to
light resonances that should have already been seen. As always we assume (rightly or wrongly) that any scalar resonance above
1.8 TeV should have been observed. And as always, we also force the vector resonances,  if present, to be heavier than
that scale.

We do not find any physical resonant state with $M_S\gtrsim 1.8$ TeV and $d_4\gtrsim 6$.

\section{Main conclusions}
We have presented a systematical analysis to set bounds on HEFT coefficients describing the low-energy regime of a high-energy theory with strong interactions.
By making use of the partial wave analysis and unitarizarion techniques herein presented, and under some mild assumptions, we have been able to set bounds on the anomalous self-interactions 
of the Higgs pointing in the same direction that the experimental results: in particular we find $d_3<2$ and $d_4<2.5$. 

The possibility of light scalar resonances, below our self-imposed limit $M_S<1.8$ TeV, remains a logical possibility to be further studied. However, preliminary studies of a
putative resonance at 650 GeV \cite{Kundu:2022bpy}  place such possibilility in a corner or parameter space and requires the concourse of several parameters\cite{Asiain:2023myt}.

In conclusion, somewhat unexpectedly, the study of possible scalar resonances in $W W$ fusion places very interesting restrictions on the space of Higgs couplings,
a region that is hard to experimentally study. We have presented here some, we believe, relevant results, but certainly this line of research deserves furthe analysis.

\begin{small}

\bibliographystyle{utphys.bst}
\bibliography{bibliography}

\end{small}

\end{document}